# High magnetocaloric efficiency of a NiFe/NiCu/CoFe/MnIr multilayer in a small magnetic field

S. N. Vdovichev[1], N. I. Polushkin[1], I. D. Rodionov[2], V. N. Prudnikov[2], J. Chang[3], and A. A. Fraerman[1]

[1] Institute for Physics of Microstructures of RAS, GSP-105, Nizhny Novgorod, Russia
[2] Lomonosov Moscow State University, Faculty of Physics, Moscow 119991, Russia
[3] Center for Spintronics, Post-Si Semiconductor Institute, Korea Institute of Science and Technology, Hwarangno 14-gil 5, Seongbuk-gu, Seoul 02792, Korea

The isothermal magnetic entropy changes ($\Delta S_M$'s) are studied in $Ni_{80}Fe_{20}/Ni_{67}Cu_{33}/Co_{90}Cu_{10}/Mn_{80}Ir_{20}$ stacks at temperatures ($T$) near the Curie point ($T_c$) of the $Ni_{67}Cu_{33}$ spacer by applying magnetic fields ($H$) in a few tens of Oersted. Such low values of $H$ were sufficient for toggling magnetic moments in the soft ferromagnetic (FM) layer ($Ni_{80}Fe_{20}$). It is found out that this switching provides the value of $\Delta S_M$, which is up to ~20 times larger than that achievable in a single $Ni_{67}Cu_{33}$ film subjected to such low $H$. Our finding holds promise to be utilized in the magnetocaloric devices that would be based on FM/PM/FM heterostructures and would operate with moderate $H$.

*Introduction*. – In materials that exhibit a strong magnetocaloric effect (MCE), the maximal magnetocaloric efficiency is basically achievable in the vicinity of phase magnetic transitions, e.g., near $T_c$ [1, 2]. The MCE, observed currently in record magnetocaloric materials, is believed to be sufficient for cooling down (or heating up) the material by applying $H$ up to several tens of kilo-oersted over several tens of thermodynamic cycles [3-6]. However, such strong fields can only be produced with bulky magnets, which are undesired to be employed in magnetic refrigeration [7]. Therefore, there is a task to seek for MCE materials, in their both bulk and thin-film forms [8-10], that could be used for magnetic cooling with moderate fields. We propose to enhance the MCE by surrounding a magnetocaloric material by higher-$T_c$ FM's, [11, 12] so that their reconfigurations in moderate $H$ could provide magnetizing/demagnetizing the PM spacer due to the effect of proximity [13]. It has been shown [12, 14-17] that FM layers surrounding the PM (or weakly FM) spacer strongly affect its magnetization up to the spacer thickness of ~ 20 nm [12].

Here we report on our measurements of $\Delta S_M$ in $Ni_{80}Fe_{20}/Ni_{67}Cu_{33}/Co_{90}Cu_{10}/Mn_{80}Ir_{20}$ stacks at $T$ close to $T_c$ of the $Ni_{67}Cu_{33}$ spacer between the FM layers, i.e., $Ni_{80}Fe_{20}$ and $Co_{90}Fe_{10}$. Such a heterostructure system exhibits reconfigurations of the mutual orientation of FM magnetizations, $\mathbf{M}_1$ and $\mathbf{M}_2$, under applying a field of $H_{sw}$~20 Oe due to their exchange decoupling across the spacer above its $T_c$ [14-17]. Our evaluations of the MCE in FM/PM/FM structures anticipate a very high MCE, $dT/dH \approx 2.0$ K/kOe in bias fields of only a few tens of oersted [11]. This MCE originates from magnetizing/demagnetizing the PM spacer when the mutual orientation of $\mathbf{M}_1$ and $\mathbf{M}_2$, alters from parallel (antiparallel) to antiparallel (parallel). Such a reconfiguration in a F$_1$/PM/F$_2$/AF stack, where F$_1$=$Ni_{80}Fe_{20}$, PM= $Ni_{67}Cu_{33}$, F$_2$= $Co_{90}Fe_{10}$, and AF=$Mn_{80}Ir_{20}$ is the antiferromagnetic layer, is schematically shown in Fig. **1**. The role of the AF layer is to render the layer F$_2$ magnetically hard, which would contrasts to the magnetically soft layer F$_1$. The dependence of $\Delta S_M$ on the mutual orientation of magnetizations of $\mathbf{M}_1$ and $M_2$ in $F_1$ and $F_2$ layers is associated with the giant MCE by analogy with the effect of giant magnetoresistance in spin valves [18, 19]. It is also important that the $T_c$ of the spacer can be tunable by varying the spacer composition. For example, a diluted $Ni_xCu_{1-x}$ alloy, whose $T_c$ depends almost linearly on the Ni concentration, is a good candidate as the spacer material [14-17].

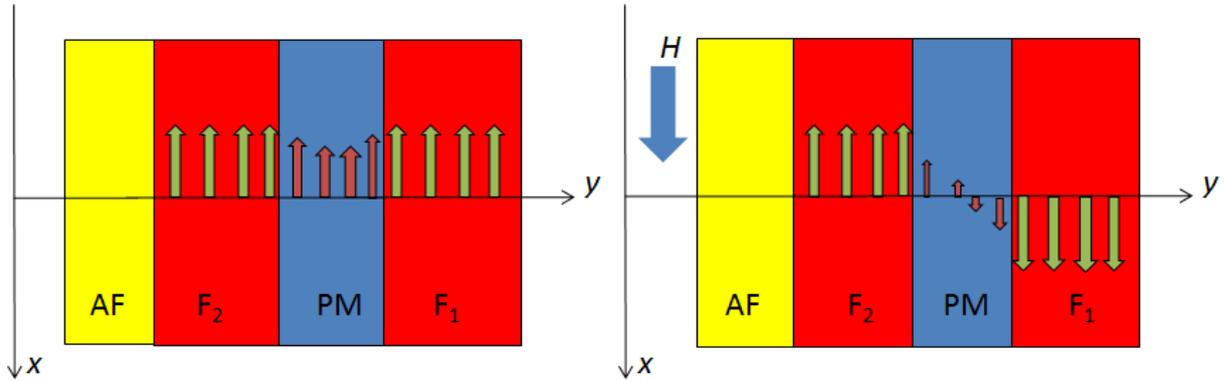

Fig.1. Schematic of the magnetization configurations in the $F_1$/PM/$F_2$/AF structure, where AF is the antiferromagnet, whose role is to pin magnetic moments in the layer $F_2$, which would be in contrast to the soft layer $F_1$. If magnetizations in the FM layers, $F_1$ and $F_2$, are oriented in the same direction (left), the spacer is magnetized (low magnetic entropy) by the exchange interactions with the $F_1$ and $F_2$ layers across the interfaces, while the spacer is demagnetized (high magnetic entropy) in the opposite case by reconfiguring magnetizations in the FM layers (right) with an external magnetic field $H$. In this thought experiment, the temperature is close to $T_c$ of the spacer. Therefore, the reconfiguration of magnetizations in the FM layers is expected to provide a strong MCE [11, 12] by analogy with the effect of giant magnetoresistance [18, 19].

*Samples preparation and characterization.* – $F_1$/PM/$F_2$/AF samples were fabricated with high-vacuum magnetron sputtering (AJA 2200 multichamber system) onto fused polished quartz substrates at a basic pressure of ~$10^{-7}$ Torr. In order to prevent them from oxidation, the samples were covered by a layer of TaO. The layers of $Ni_{80}Fe_{20}$ ($F_1$), $Co_{90}Fe_{10}$ ($F_2$), and $Mn_{80}Ir_{20}$ (AF) were sputtered from single targets, while PM films of $Ni_xCu_{1-x}$ solid solutions were obtained by simultaneous sputtering of Ni and Cu targets and their composition was determined by X-ray microanalysis using a dispersion spectrometer (INCA Energy Oxford Instruments). The exchange pinning between the $F_2$ and AF layers was achieved during the film growth with assisting an in-plane magnetic field of $H\approx0.2$ kOe. Sample magnetization as a function of $H$ was measured with a Lake Shore 7400 Series vibrating sample magnetometer and a MPMS-XL5 Quantum Design SQUID magnetometer in a 5–400 K temperature range and external magnetic fields up to $H=5$ kOe applied in the film plane. The sample area used in these studies was $S=0.2$ cm$^2$. The individual magnetizations of the FM layers, $M_1$ and $M_2$, were determined from the Kittel equation by measuring the ferromagnetic resonance [12]. The composition of the $Ni_xCu_{1-x}$ solid solution was $x=67$ at. %, while the thickness of the $Ni_{67}Cu_{33}$ spacer was varied from 7 nm to 21 nm. The thickness of the soft $Ni_{80}Fe_{20}$ layer ($F_1$) was $h_{NiFe}=(10\pm1)$ nm and these quantities for the rest of the layers were as follows: 5 nm of $Co_{90}Fe_{10}$, 25 nm of $Mn_{80}Ir_{20}$, and 15 nm of TaO. The layer thicknesses were determined by small angle X-ray diffraction (reflectometry Bruker D8 Discover).

*Results and discussion.* - Figure 2 (a) shows the magnetization ($M$) isotherms, measured as the total magnetic moment $\mu\equiv MV$ of a TaO/$Mn_{80}Ir_{20}$/$Co_{90}Fe_{10}$/(10 nm)$Ni_{67}Cu_{33}$/$Ni_{80}Fe_{20}$/substrate

sample with a $Ni_{67}Cu_{33}$ thickness of 10 nm as a function of $H$, at two temperatures, $T$=250 K and $T$=300 K. In the observed $M(H)$ dependences we distinguish two steps which correspond to switching the soft and pinned layers. Importantly, the observed change in $MV$ at the first (lower-$H$) step, that results from the switching of magnetization ($M_1$) in the soft $Ni_{80}Fe_{20}$ layer ($F_1$), is $\Delta\mu$=3.2×10$^{-4}$ emu. This magnitude quantitatively agrees with the value of $M_1$=730 emu/cm$^3$ [12], so that $\Delta\mu=2M_1 Sh_{NiFe}$. The observed delay in the switching of pinned layer of $Co_{90}Fe_{10}$ ($F_2$) results from the exchange coupling of magnetic moments across the $F_2$/AF interface. We also see that the field $H_{sw}$, at which the soft layer is switchable, increases with lowering $T$, so its reversal becomes undistinguishable from that of the pinned layer at low enough $T$<100 K [14]. In Fig. **2 (b)** we show $H_{sw}$ versus $T$ near $T_c\approx$240 K of the $Ni_{67}Cu_{33}$ [20] for two samples – with thicknesses of the $Ni_{67}Cu_{33}$ spacer of 10 nm and 21 nm. The fact that $H_{sw}$ depends on $T$ indicates that the interaction through the spacer is sensitive to the temperature. We see that the temperature derivative of the effective interlayer exchange constant $J'(T)$=(1/$M_2$)$dH_{sw}/dT$ [see **Appendix**] increases with lowering $T$ below $T_c$. We do not mark a large difference between $J'(T)$ for the samples with different thicknesses of the spacers.

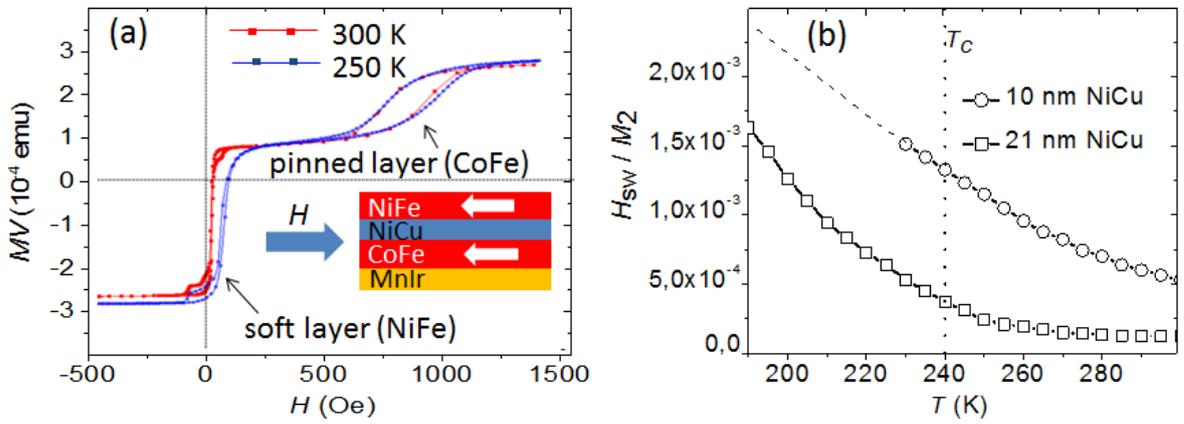

Fig.**2**. (**a**) Isothermal magnetization of a TaO/$Mn_{80}Ir_{20}$/$Co_{90}Fe_{10}$/(10 nm)$Ni_{67}Cu_{33}$/$Ni_{80}Fe_{20}$/substrate sample at $T$=300 K and $T$=250 K. In the response to the applied magnetic field $H$, the magnetization reversal occurs in two steps that correspond to switching the soft $Ni_{80}Fe_{20}$ layer and pinned $Co_{90}Fe_{10}$ layer. The reversal of **M**$_2$ in the pinned layer is delayed due to exchange coupling of this layer with the antiferromagnetic layer of $Mn_{80}Ir_{20}$. The inset illustrates the mutual configuration of an applied field **H** and FM magnetizations, **M**$_1$ and **M**$_2$, before switching magnetic moments in the soft layer (NiFe), which occurs at $H$=$H_{sw}$. (**b**) The switching field $H_{sw}$ measured and normalized by $M_2$ as a function of $T$ for the 10- and 21-nm-thick $Ni_{67}Cu_{33}$ spacers. Accordingly to Eq. (2) presented in **Appendix**, this quantity is interpretable as the interlayer exchange constant $J$.

Figure **3** (**a**) shows $M(T)$ dependences measured for the sample with the 10-nm-thick $Ni_{67}Cu_{33}$ spacer in a field range of 8÷30 Oe with a step in $H$ of 2 Oe. These data were collected by applying $H$ against magnetizations of **M**$_1$ and **M**$_2$ in the configuration of their parallel orientation with respect to each other. We have found that the response of $M$ strongly depends on both $T$ and $H$, especially in a temperature range of 200÷280 K at $H$~20 Oe. The experimental data plotted in Fig. **3 (a)** allow us to account $\mu_0(\partial M/\partial T)_H$–versus–$H$ curves, where $\mu_0$=4$\pi$×10$^{-7}$ J/A$^2$m is the magnetic constant. These curves are shown in Fig. **3 (b)** for several values of $T$. According to Eq. (1), areas under these curves are the $\Delta S_M$'s. In Fig. **3 (c)** we compare the peak values of $\mu_0(\partial M/\partial T)_H$ for the samples with the 10- and 21-nm-thick spacers at $T$=250 K and $T$=210 K, respectively, which provide the largest values of $\Delta S_M$. Figure **3 (d)** shows $\Delta S_M(T)$ dependences for the samples with the 7-, 10-, and 21-nm-thick $Ni_{67}Cu_{33}$ spacers. The obtained values of $\Delta S_M$ can be compared to those quantities in a single $Ni_{67}Cu_{33}$ film

subjected to a magnetic field of $H=20$ Oe. The values of $\Delta S_M$ in low $H$ for a $Ni_{67}Cu_{33}$ film were retrieved by extrapolation into the low-field region of the $\Delta S_M(H) \propto H^n$ dependences [21] with $n \approx 0.8$ at $T$ close to $T_c \approx 240$ K of the $Ni_{67}Cu_{33}$, which have been reported in Ref. [20]. Although the maximal peak values of $\mu_0(\partial M/\partial T)_H$ are found to be highest for the sample with a thinner (10 nm) spacer (Fig. **3c**), we mark that the maximal $\Delta S_M \approx 120$ µJ/cm³K is in the sample with a thicker (21 nm) spacer (Fig. **3d**). Strikingly, the maximal quantity of $\Delta S_M$ obtained in the heterostructure systems is nearly 20 times larger than that in a single $Ni_{67}Cu_{33}$ film and should provide cooling of the refrigerant by $\Delta T_{ad}=(T/C_H)\Delta S_M \approx 0.008°$ provided that the heat capacity of our sample is $C_H=3.5$ J/cm³K [22]. Note here that the magnetocaloric efficiency in the heterostructure system under study is $\partial T/\partial H=-(T/C_H)\mu_0(\partial M/\partial T) \approx -0.7$ K/kOe, which is comparable to this quantity in advanced magnetocaloric materials. [**1**, **2**]

It is also important that the $\Delta S_M$ can alternatively be evaluated from Eq. (4) derived in **Appendix** after measuring $H_{sw}(T)$ (Fig. **2 b**) and subsequently finding $J(T)$ according to Eq. (2) in **Appendix**. We find that the maximal $J'(T)=4.0\times10^{-5}$ K$^{-1}$ is achievable for a sample with the thickest (21 nm) $Ni_{67}Cu_{33}$ spacer at $T \approx 210$ K (Fig. **2b**). As the FM magnetizations are found to be $M_1=0.73\times10^6$ Am$^{-1}$ and $M_2=1.5\times10^6$ Am$^{-1}$, [**12**] we obtain that $\Delta S_M=114$ µJ/Kcm³, which well agrees with that straightforwardly calculated from the experimental data, using the Maxwell relation. Note also that both ways give similar $\Delta S_M(T)$ dependences and provide the same ratios between $\Delta S_M$'s, obtained for the samples with the spacers of a different thickness.

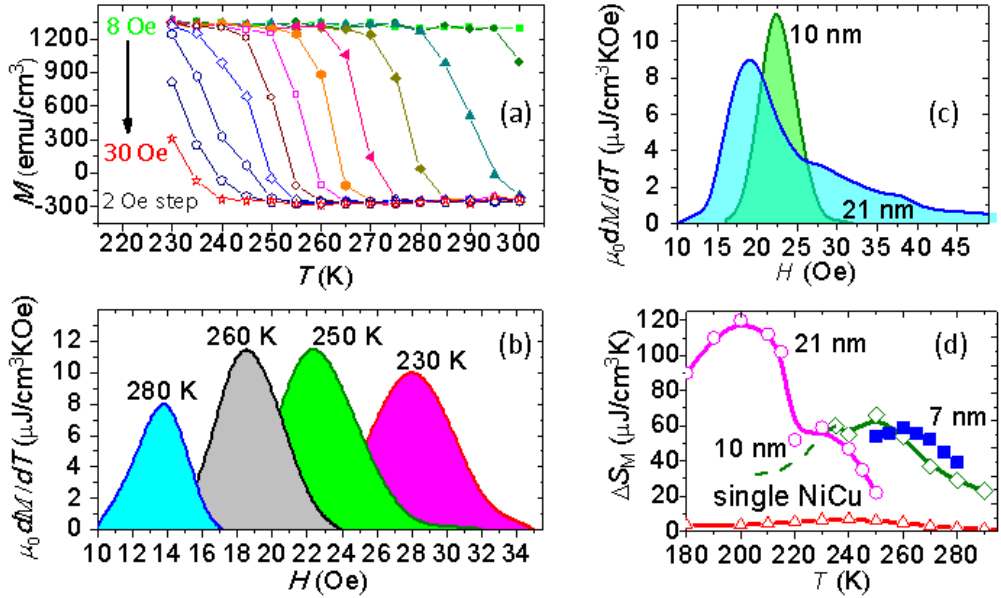

Fig.**3**. (**a**) $M(T)$ dependences measured in a field range of 8÷30 Oe with a step of 2 Oe for the sample with the 10-nm-thick $Ni_{67}Cu_{33}$ spacer. (**b**) The $\mu_0 \partial M/\partial T$–versus–$H$ curves at several $T$ close to T$_c$ of the $Ni_{67}Cu_{33}$ spacer calculated for the same sample – with the 10-nm-thick spacer, whose $M(T)$ dependences are shown in (**a**). Areas under the curves are $\Delta S_M$. (**c**) The $\mu_0 \partial M/\partial T$–versus–$H$ curves for the samples with the 10- and 21-nm thick spacers at $T=250$ K and $T=210$ K, respectively, which provide the largest values of $\Delta S_M$. (**d**) $\Delta S_M(T)$ curves evaluated by integration accordingly to Eq. (1) for the samples with 10-, 21-, and 7-nm-thick $Ni_{67}Cu_{33}$ spacers. The obtained values of $\Delta S_M$ for the heterostructure samples are compared to that quantity in a single $Ni_{67}Cu_{33}$ film, which is retrievable by extrapolation into the low-field region ($H \sim 20$ Oe) of the experimental $\Delta S_M(T)$ dependences obtained in Ref. [**20**] in high $H \geq 10^4$ Oe.

In order to qualitatively explain the obtained gain in $\Delta S_M$ in the $F_1$/PM/$F_2$/AF system, we schematically (Fig. **4**) consider the isothermal magnetization of this system as a function of $H$ for two temperatures $T_1$ and $T_2$, so that $T_1>T_2$ and thus $J(T_1)<J(T_2)$. If, for instance, the system stays in the state A at $T_1$, lowering the temperature to $T_2$ provides a transition of the system to the state B – with a strongly different magnetization, as indicated by the arrow. This transition results from the switching of magnetic moments in the soft layer ($F_1$). Importantly, the total magnetic moment alters by the magnitude of the magnetic moment of the layer of $F_1$. This gives a high magnetocaloric efficiency observed in our experiments. A temperature range, within of which the A→B transition occurs, is defined by the efficiency of the interlayer exchange as a function of $T$, i.e., by $J'(T)$. The change of the magnetic moment under varying $T$ occurs within a finite field range, $J(T_1)M_2<H_A<J(T_2)M_2$. With increasing a biasing field $H_A>H_{sw}\sim 20$ Oe, $\partial M/\partial T$ quickly decays and can be neglected at $H_A>100$ Oe (Fig. **3b**). Decreasing a biasing field, so that $H_A<H_{sw}$, also leads to decreasing the MCE. A quick decay of $\partial M/\partial T$ imposes a constraint for the adiabatic temperature change ($\Delta T_{ad}$) per one magnetization/demagnetization cycle. In order to achieve a large $\Delta T_{ad}$, many of refrigeration cycles are needed, which can be employed for regenerative cooling.

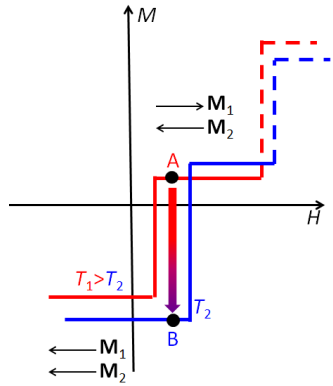

Fig.**4**. Schematic of the $M(H)$ dependence for an $F_1$/PM/$F_2$/AF heterostructure system at two temperatures. The transition of the system from the state A (antiparallel orientation of **M**$_1$ and **M**$_2$) to the state B (parallel orientation of **M**$_1$ and **M**$_2$), which is triggered by lowering the temperature from $T_1$ to $T_2$, leads to the isothermal magnetic entropy change $\Delta S_M$.

In conclusion, we show that, at temperatures close to $T_c$ of the $Ni_{67}Cu_{33}$ spacer, $Ni_{80}Fe_{20}/Ni_{67}Cu_{33}/Co_{90}Cu_{10}/Mn_{80}Ir_{20}$ multilayers exhibit the isothermal magnetic entropy changes $\Delta S_M$´s, which are much larger than this quantity in a single $Ni_{67}Cu_{33}$ paramagnet subjected to a moderate magnetic field $H$. The enhanced magnetocaloric efficiency in our samples results from the magnetic-field-driven reconfiguration of the soft ferromagnet (NiFe) with respect to the pinned one (CoFe). Switching of magnetic moments in the NiFe alters the magnetization distribution in the $Ni_{67}Cu_{33}$ spacer [**14-17**], which should provide the enhanced MCE in accordance to our calculations performed in Ref. [**11**]: The thinner spacer, the stronger its magnetization/demagnetization by the ferromagnetic surroundings, and the higher MCE. Indeed, having compared the quantities of $\mu_0\partial M/\partial T$ in the samples with thin (10 nm) and thick (21 nm) spacers, we mark that the peak values of $\mu_0\partial M/\partial T$ are higher in the sample with a thinner spacer (Fig. **3c**). Nevertheless, the quantity of $\Delta S_M$ is found to be higher in the sample with a thick spacer (Fig. **3d**). Therefore, the question about the mechanism behind the observed enhancement of the MCE in the heterostructure systems we study requires further clarifications. It is, however, important that the experimental data obtained can be quantitatively explained in a frame of the model for switching a ferromagnet in a temperature-dependent effective field $\mathbf{H}_{eff}=J(T)\mathbf{M}_2+\mathbf{H}$ (see **Appendix**). The devices upon that basis can find the applications in electronics as thin-film coolers [**3-6**] – with up to several tens of refrigeration cycles.

*Acknowledgments*. – The authors thank to P. Yunin and S. Gusev for their assistance in studying the structural properties of the samples. Work was supported by the Russian Foundation for Basic Research, project no. 17-0200620_a. J. Chang acknowledges the support of the KIST Institutional Program (2E27140)

**APPENDIX**

We wish to validate the usage of the thermodynamic Maxwell relation for accounting the isothermal magnetic entropy change,

$$\Delta S_M = \mu_0 \int_{H_{sw}-\varepsilon_1}^{H_{sw}+\varepsilon_2} (\partial M/\partial T)_H \, dH, \tag{1}$$

in the case of a nonuniform magnetic system. The notations in Eq. (1) are as follows: $M$ is the total magnetization of the system, while $\varepsilon_{1(2)}$ are so large field deviations from $H_{sw}$ that the contributions of $\partial M/\partial T$ at the integration ends, $H_{sw}-\varepsilon_1$ and $H_{sw}+\varepsilon_2$, into Eq. (1) can be discarded. Magnetizations $M_1$ and $M_2$, that largely contribute to $M$, are assumed to be coupled through the spacer by an exchange interaction of the form of $-J\mathbf{M}_1\mathbf{M}_2$ [23]. It was shown previously [14-17] that $J$ is sensitive to the temperature, especially in the vicinity of $T_c$ of the spacer. An external magnetic field $H$ is applied in the film plane and aligned to the easy axis of the uniaxial magnetic anisotropy, $K$. Differentiating the free-energy density $f/\mu_0=-J(T)\mathbf{M}_1\mathbf{M}_2-K(\mathbf{nM}_1)^2/2-\mathbf{HM}_1$ on $\mathbf{M}_1$, where $\mathbf{n}$ is the unit vector along the easy axis, one gets the effective field acting on $\mathbf{M}_1$, i.e., $\mathbf{H}_{eff}=\mathbf{H}+K\mathbf{M}_1+J(T)\mathbf{M}_2$. At $H=H_{sw}$, we have that $H_{eff}=0$. Therefore, with neglecting the contribution of the magnetic anisotropy ($K=0$), the interlayer exchange constant can be retrieved from the following relation:

$$J(T)=H_{sw}(T)/M_2. \tag{2}$$

In terms of the effective field, the equation for the free-energy density can be written as $f=-\mu_0 H_{eff} M$, where

$$M=M_1\theta[H_{eff}(T, H)]-M_2 \tag{3}$$

with $\theta(H_{eff})=\pm 1$ at $H_{eff}>0$ (upper sign) and $H_{eff}<0$ (lower sign). After differentiating $\partial S_M/\partial(\mu_0 H)=-\partial^2 f/\partial T\partial(\mu_0 H)$, taking into account that $-(dJ/dT)M_1M_2 d\theta/dH_{eff}=\partial M/\partial T$, and subsequent integration on $H$ between $H=H_{sw}+\varepsilon_2$ and $H=H_{sw}-\varepsilon_1$ one gets Eq. (1). Upon substituting Eq. (3) into Eq.(1) and its subsequent integration, we have finally that

$$\Delta S_M = -2\mu_0 J'(T) M_1 M_2, \tag{4}$$

where $J'(T) \equiv dJ/dT$.